\documentclass{fcs}
\pdfoutput=1
\usepackage[pdftex,dvipsnames]{xcolor}
\usepackage[letterpaper,margin=1in,footskip=0.25in]{geometry}
\usepackage{bm}
\usepackage{tikz}
\usepackage{listings}
\usepackage{subfigure}
\usepackage[colorinlistoftodos,prependcaption]{todonotes}
\usepackage{amsmath,amssymb}

\usetikzlibrary{shapes%
	,positioning %
	,arrows %
	,calc%
} 

\newcommand{\overbar}[1]{\mkern 1.5mu\overline{\mkern-1.5mu#1\mkern-1.5mu}\mkern 1.5mu}
\newcommand{\nop}[1]{}
\newcommand{\shiftAndRotate}[4]{
	\coordinate (#1) at ($(#2) + ( {#3*cos(#4)}, {#3*sin(#4)} )$);
}
\newcommand{\drawTriple}[6]{
	\node[draw, circle, fill={rgb:black,1;white,12}, minimum size=1.25cm, inner sep=1pt, name=vertex-#1] at (#1) {\large $\mathbf{ #3 }$};
	\node[draw, circle, fill={rgb:black,1;white,12}, minimum size=1.25cm, inner sep=1pt, name=vertex-#2] at (#2) {\large $\mathbf{ #4 }$};
	\path[draw=black,solid,line width=0.25mm,fill=black,-triangle 60] (vertex-#1) -- (vertex-#2)  node[midway,#6,color=blue] {\large $\mathbf{ #5 }$};
}

\tikzstyle{every entity} = []
\tikzstyle{every attribute} = []
\tikzstyle{every link} = []

\tikzstyle{link} = [>=triangle 60, draw, thick, every link]

\tikzstyle{entity} = [rectangle, draw, black, very thick,
                      minimum width=6em, minimum height=3em,
                      every entity]

\tikzstyle{smentity} = [rectangle, draw, black, 
                      minimum width=4em, minimum height=2em,
                      every entity]

\tikzstyle{smentityd} = [rectangle, draw, dashed, black, 
                      minimum width=4em, minimum height=2em,
                      every entity]

\tikzstyle{attribute} = [rectangle, 
                         minimum width=4em, minimum height=2em,
                         every attribute]

\volumn{ }
\doi{ }
\articletype{RESEARCH~ARTICLE}
\copynote{{\copyright} Higher Education Press and Springer-Verlag Berlin Heidelberg 2012}
\ratime{Received month dd, yyyy; accepted month dd, yyyy}
\email{tamer.ozsu@uwaterloo.ca}
\title{A Survey of RDF Data Management Systems}
\author{M. Tamer \MakeUppercase{\"Ozsu}\xff}
\address{{University of Waterloo, Cheriton School of Computer Science, Canada}}

\begin{document}
\maketitle
\setcounter{page}{1}
\setlength{\baselineskip}{14pt}

\begin{abstract}
RDF is increasingly being used to encode data for the semantic web and for data exchange. There have been a large number of works that address RDF data management. In this paper we provide an overview of these works.

\end{abstract}


\section{Introduction}
\label{sec:intro}

The RDF (\textbf{R}esource \textbf{D}escription \textbf{F}ramework) is a W3C standard that was proposed for modeling Web objects as part of developing the semantic web. However, its use is now wider than the semantic web. For example, Yago and DBPedia extract facts from Wikipedia automatically and store them in RDF format to support structural queries over Wikipedia\cite{Fabian:yagoWWW07,Bizer:dbpedia}; biologists  encode their experiments and results using RDF to communicate among themselves leading to RDF data collections, such as Bio2RDF ({bio2rdf.org}) and Uniprot RDF (dev.isb-sib.ch/projects/ uniprot-rdf). Related to semantic web, LOD (Linking Open Data) project builds a RDF data cloud by linking more than 3000 datasets, which currently have more than 84 billion triples\footnote{The statistic is reported in http://stats.lod2.eu/.}. A recent work \cite{SBP14} shows that the number of data sources in LOD has doubled within three years (2011-2014).

RDF data sets have all four accepted characteristics of ``big data'': volume, variety, velocity, and veracity. We hinted at increasing volumes above. RDF data, as captured in the LOD cloud, is highly varied with many different types of data from very many sources. Although early (and still much of) RDF data sets are stationary, there is increasing interest in streaming RDF to the extent that W3C has now set up a community interested in addressing problems of high velocity streaming data (www.w3.org/community/rsp/). RDF-encoded social network graphs that continually change is one example of streaming RDF. Even a benchmark has been proposed for this purpose~\cite{Zhang:2012aa}. Finally, it is commonly accepted that RDF data is ``dirty'', meaning that it would contain inconsistencies. This is primarily as a result of automatic extraction and conversion of data into RDF format; it is also a function of the fact that, in the semantic web context, the same data are contributed by different sources with different understandings. A line of research focuses on data quality and cleaning of LOD data and RDF data in general~\cite{Zaveri:2012aa,Tang:2015aa}.

Because of these characteristics, RDF data management has been receiving considerable attention. In particular, the existence of a standard query language, SPARQL, as defined by W3C, has given impetus to works that focus on efficiently executing SPARQL queries over large RDF data sets. In this paper, we provide an overview of research efforts in management of RDF data.  We note that the amount of work in this area is considerable and it is not our aim to review all of it. We hope to highlight the main approaches and refer to some of the literature as examples. Furthermore, due to space considerations, we limit our focus on the management of stationary RDF data, ignoring works on other aspects. 

The organization of the paper is as follows. In the next section (Section \ref{sec:primer}) we provide a high level overview of RDF and SPARQL to establish the framework. In Section \ref{sec:warehouse}, the centralized approaches to RDF data management are discussed, whereas Section \ref{sec:distributed} is devoted to a discussion of distributed RDF data management. Efforts in querying the LOD cloud is the subject of Section \ref{sec:lod}.

\section{RDF Primer}
\label{sec:primer}

In this section, we provide an overview of RDF and SPARQL. The objective of this presentation is not to cover RDF fully, but to establish a basic understanding that will assist in the remainder of the paper. For a fuller treatment, we refer the reader to original sources of RDF~\cite{Klyne:aa,Harris:aa}. This discussion is based on \cite{Zou:2013fk} and \cite{HartigOzsu2014}.

RDF models each ``fact'' as a set of triples (\textbf{s}ubject, \textbf{p}roperty (or \textbf{p}redicate), \textbf{o}bject), denoted as $\langle s, p, o \rangle$, where \emph{subject} is an entity, class or blank node, a \emph{property}\footnote{In literature, the terms ``property'' and ``predicate'' are used interchangeably; in this paper, we will use ``property'' consistently.} denotes one attribute associated with one entity, and \emph{object} is an entity, a class, a blank node, or a literal value. According to the RDF standard, an entity is denoted by a URI (Uniform Resource Identifier) that refers to a named \emph{resource} in the environment that is being modelled. Blank nodes, by contrast,  refer to  anonymous resources that do not have a name\footnote{In much of the research, blank nodes are ignored. Unless explicitly stated otherwise, we will ignore them in this paper as well.}. Thus, each triple represents a named relationship; those involving blank nodes simply indicate that ``something with the given relationship exists, without naming it'' \cite{Klyne:aa}.

It is possible to annotate RDF data with semantic metadata using RDFS (RDF Schema) or OWL, both of which are W3C standards. This annotation primarily enables reasoning over the RDF data (called entailment), that we do not consider in this paper. However, as we will see below, it also impacts data organization in some cases, and the metadata can be used for semantic query optimization. We illustrate the fundamental concepts by simple examples using RDFS, which allows the definition of \emph{classes} and \emph{class hierarchies}. RDFS has built-in class definitions -- the more important ones being {\small \textsf{rdfs:Class}} and  {\small \textsf{rdfs:subClassOf}} that are used to define a class and a subclass, respectively (another one,  {\small \textsf{rdfs:label}} is used in our query examples below). To specify that an individual resource is an element of the class, a special property,  {\small \textsf{rdf:type}} is used. For example, if we wanted to define a class called  {\small \textsf{Movies}} and two subclasses  {\small \textsf{ActionMovies}} and  {\small \textsf{Dramas}}, this would be accomplished in the following way: 

\noindent
 {\small 
\textsf{
Movies rdf:type rdfs:Class . \\
ActionMovies rdfs:subClassOf Movies . \\
Dramas rdfs:subClassOf Movies .
}}

\begin{definition}[RDF data set]\label{def:rdf}
Let $\mathcal{U}, \mathcal{B}, \mathcal{L},$ and $\mathcal{V}$ denote the sets of all URIs, blank nodes, literals, and variables, respectively. A tuple $(s,p,o) \in (\mathcal{U} \cup \mathcal{B}) \times \mathcal{U} \times (\mathcal{U} \cup \mathcal{B} \cup \mathcal{L})$ is an \textit{RDF triple}. A set of RDF triples form a \textit{RDF data set}.

\end{definition}

An example RDF data set  is shown in Figure \ref{fig:rdftable} where the data comes from a number of sources as defined by the URI prefixes.

\begin{figure}
	\scriptsize
	Prefixes: \\
	mdb=http://data.linkedmdb.org/resource/
	geo=http://sws.geonames.org/ \\
	bm=http://wifo5-03.informatik.uni-mannheim.de/bookmashup/ \\  
	exvo=http://lexvo.org/id/ \\
	wp=http://en.wikipedia.org/wiki/
	
	\begin{tabular}{|l|l|l|c|} 
		\hline
		Subject & Property & Object \\ \hline
		mdb: film/2014 & rdfs:label & ``The Shining'' \\
		mdb:film/2014 & movie:initial\_release\_date &  ``1980-05-23''' \\
		mdb:film/2014 & movie:director & mdb:director/8476 \\
		mdb:film/2014  &  movie:actor & mdb:actor/29704 \\
		mdb:film/2014 & movie:actor & mdb: actor/30013\\
		mdb:film/2014  &  movie:music\_contributor & mdb: music\_contributor/4110 \\
		mdb:film/2014  &  foaf:based\_near & geo:2635167 \\
		mdb:film/2014  &  movie:relatedBook & bm:0743424425 \\
		mdb:film/2014  &  movie:language & lexvo:iso639-3/eng \\
		mdb:director/8476 & movie:director\_name & ``Stanley Kubrick'' \\
		mdb:film/2685   &  movie:director & mdb:director/8476 \\
		mdb:film/2685  &  rdfs:label & ``A Clockwork Orange'' \\
		mdb:film/424  &  movie:director & mdb:director/8476 \\
		mdb:film/424   &  rdfs:label & ``Spartacus'' \\
		mdb:actor/29704  &  movie:actor\_name & ``Jack Nicholson'' \\
		mdb:film/1267  &  movie:actor & mdb:actor/29704 \\
		mdb:film/1267  &  rdfs:label & ``The Last Tycoon'' \\
		mdb:film/3418  &  movie:actor & mdb:actor/29704 \\
		mdb:film/3418  &  rdfs:label & ``The Passenger'' \\
		geo:2635167  &  gn:name & ``United Kingdom'' \\
		geo:2635167  &  gn:population & 62348447 \\
		geo:2635167  &  gn:wikipediaArticle & wp:United\_Kingdom \\
		bm:books/0743424425  &  dc:creator & bm:persons/Stephen+King \\
		bm:books/0743424425 & rev:rating  &   4.7 \\
		bm:books/0743424425 & scom:hasOffer & bm:offers/0743424425amazonOffer \\
		lexvo:iso639-3/eng & rdfs:label   & ``English'' \\
		lexvo:iso639-3/eng & lvont:usedIn  &    lexvo:iso3166/CA \\
		lexvo:iso639-3/eng & lvont:usesScript & lexvo:script/Latn \\
		\hline
	\end{tabular}
	\caption{Example RDF dataset. Prefixes are used to identify the data sources.}
	\label{fig:rdftable}
\end{figure}

RDF data can  be modeled as an RDF graph, which is formally defined as follows. 

\begin{definition}[RDF graph]\label{def:rdfgraph}
A \emph{RDF graph} is a six-tuple $G=\langle V,L_V, f_{V}, E,L_E, f_{E} \rangle$, where

\begin{enumerate}
	\item $V=V_c\cup V_e\cup V_l$ is a collection of vertices that correspond to all subjects and objects in RDF data, where $V_c$, $V_e$, and $V_l$ are collections of class vertices, entity vertices, and literal vertices, respectively.
	\item $L_V$ is a collection of vertex labels. 
	\item A \emph{vertex labeling function} $f_{V}: V \rightarrow L_{V}$ is an bijective function that assigns to each vertex a label. The label of a vertex $u \in V_l$ is its literal value, and the label of a vertex $u \in V_c \cup V_e$ is its corresponding URI.
	\item $E=\{\overrightarrow{u_1,u_2}\}$ is a collection of directed edges that connect the corresponding subjects and objects.
	\item $L_E$ is a collection of edge labels. 
	\item An \emph{edge labeling function} $f_{E}: E \rightarrow L_{E}$ is an bijective function that assigns to each edge a label. The label of an edge  $e \in E$ is its corresponding property.
	
\end{enumerate}
An edge $\overrightarrow{u_1,u_2}$ is an \emph{attribute property}  edge if $u_2 \in V_l$; otherwise, it is a \emph{link} edge.

\end{definition}

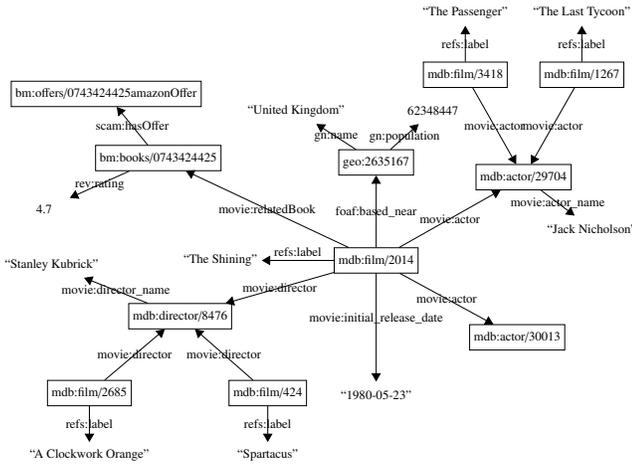
\begin{figure}[h]
\scalebox{0.48}{
	\begin{tikzpicture}[node distance=1.5cm, every edge/.style={link}]

%
%
  \node[smentity] (001) {mdb:film/2014};
  \node[attribute] (r1a1) [below =3cm  of 001] {``1980-05-23''} edge [<-] node[above] {movie:initial\_release\_date} (001);
  \node[attribute] (r1a2) [left =2cm  of 001] {``The Shining''} edge [<-] node[above] {refs:label} (001);

%
%
  \node[smentity] (002) [above left=3cm of 001,xshift=-1cm] {bm:books/0743424425};
  \node[attribute] (r2a1) [below left = 1cm of 002] {4.7} edge [<-] node {rev:rating} (002);
%
%
  \node[smentity] (003) [above left = 1.5cm of 002,xshift=4cm] {bm:offers/0743424425amazonOffer};
%
%
  \node[smentity] (004)[above = 2cm  of 001]{geo:2635167};
  \node[attribute] (r4a1) [above left = 1cm of 004,xshift=1cm] {``United Kingdom''} edge [<-] node {gn:name} (004);
  \node[attribute] (r4a2) [above right = 1cm of 004,xshift=-1cm] {62348447} edge [<-] node {gn:population} (004);
%
%
  \node[smentity] (005)[above right= 2.25cm of 001]{mdb:actor/29704};
  \node[attribute] (r5a1) [below right= 1cm of 005,xshift=-1.5cm] {``Jack Nicholson''} edge [<-] node {movie:actor\_name} (005);
%
%
  \node[smentity] (006)[above left=3cm of 005,xshift=3cm]{mdb:film/3418};
  \node[attribute] (r6a1) [above=1cm of 006] {``The Passenger''} edge [<-] node {refs:label} (006);
%
%
  \node[smentity] (007)[above right=3cm of 005,xshift=-3cm]{mdb:film/1267};
  \node[attribute] (r7a1) [above =1cm of 007] {``The Last Tycoon''} edge [<-] node {refs:label} (007);
%
%
  \node[smentity] (008)[below left=4cm of 001,yshift=2cm]{mdb:director/8476};
  \node[attribute] (r8a1) [above left =1cm of 008] {``Stanley Kubrick''} edge [<-] node {movie:director\_name} (008);
%
  \node[smentity] (009)[below left=2cm of 008,xshift=1.5cm]{mdb:film/2685};
  \node[attribute] (r9a1) [below =1cm of 009] {``A Clockwork Orange''} edge [<-] node {refs:label} (009);
%
%
  \node[smentity] (010)[below right=2cm of 008,xshift=-1.5cm]{mdb:film/424};
  \node[attribute] (r10a1) [below =1cm of 010] {``Spartacus''} edge [<-] node {refs:label} (010);
%
%
  \node[smentity] (011)[below right=2cm of 001]{mdb:actor/30013};

%
%
   \draw[link,->] (001) to node {movie:relatedBook} (002);
   \draw[link,->] (002) to node {scam:hasOffer} (003);
   \draw[link,->] (001) to node {foaf:based\_near} (004);
   \draw[link,->] (001) to node {movie:actor} (005);
   \draw[link,->] (001) to node {movie:director} (008);
   \draw[link,->] (001) to node {movie:actor} (011);
   \draw[link,->] (006) to node {movie:actor} (005);
   \draw[link,->] (007) to node {movie:actor} (005);
   \draw[link,->] (009) to node {movie:director} (008);
   \draw[link,->] (010) to node {movie:director} (008);
\end{tikzpicture}
}
	\caption{RDF graph corresponding to the dataset in Figure \ref{fig:rdftable}}  
	\label{fig:datagraph}
\end{figure}

Figure \ref{fig:datagraph} shows an example of an RDF graph. The vertices that are denoted by boxes are entity or class vertices, and the others are literal vertices.

The W3C standard language for RDF is SPARQL, which can be defined as follows \cite{esws12_Hartig12} (for a more formal definition, we refer to the W3C specification \cite{W3C:2006aa}).

\begin{definition}[SPARQL query]\label{def:sparql}
Let $\mathcal{U}, \mathcal{B}, \mathcal{L},$ and $\mathcal{V}$ denote the sets of all URIs, blank nodes, literals, and variables, respectively. A SPARQL expression is expressed recursively

\begin{itemize}
	\item A \emph{triple pattern}  $ (\mathcal{U} \cup \mathcal{B} \cup \mathcal{V} ) \times (\mathcal{U} \cup \mathcal{V} ) \times (\mathcal{U} \cup \mathcal{B} \cup \mathcal{L} \cup \mathcal{V} )$ is a SPARQL expression,
	\item (optionally) If $P$ is a SPARQL expression, then $P~FILTER~R$ is also a SPARQL expression where $R$ is a built-in SPARQL filter condition,
	\item (optionally) If $P_{1}$ and $P_{2}$ are  SPARQL expressions, then $P_{1}~AND | OPT | OR~P_{2}$ are also SPARQL expressions.
\end{itemize}
\end{definition}

A set of triple patterns is called \emph{basic graph pattern} (BGP) and SPARQL expressions that only contain these are called \emph{BGP queries}. These are the subject of most of the research in SPARQL query evaluation.

An example SPARQL query that finds the names of the movies directed by ``Stanley Kubrick'' and have a related book that has a rating greater than 4.0 is specified as follows:

\begin{lstlisting}[language=SPARQL,basicstyle=\small,showstringspaces=false]
SELECT ?name
WHERE {
  ?m rdfs:label ?name. ?m movie:director ?d. 
  ?d movie:director_name "Stanley Kubrick".
  ?m movie:relatedBook ?b. ?b rev:rating ?r.
  FILTER(?r > 4.0) 
}
\end{lstlisting}

In this query, the first three lines in the WHERE clause form a BGP consisting of five triple patterns. All triple patterns in this example have \emph{variables}, such as ``?m'', ``?name'' and ``?r'', and ``?r'' has a filter: FILTER(?r > 4.0).

A SPARQL query can also be represented as a \emph{query graph}:

\begin{definition}[SPARQL query graph]\label{def:querygraph}
	A \emph{query graph} is a seven-tuple $Q=\langle V^Q,$ $L^Q_V,E^Q,L^Q_E, f_{V}^{Q},f_{E}^{Q},FL \rangle$, where

\begin{enumerate}
\item $V^Q=V^Q_c\cup V^Q_e\cup V^Q_l\cup V^Q_p$ is a collection of vertices that correspond to all subjects and objects in a SPARQL query, where $V^Q_p$ is a collection of variable vertices (corresponding to variables in the query expression), and $V^Q_c$ and $V^Q_e$ and $V^Q_l$ are collections of class vertices, entity vertices, and literal vertices in the query graph $Q$, respectively.
\item $E^Q$ is a collection of edges that correspond to properties in a SPARQL query. 
\item $L^Q_V$ is a collection of vertex labels in $Q$ and $L_E^Q$ is the edge labels in $E^Q$. 
\item $f_{V}^{Q}:V^{Q} \rightarrow L_{V}^{Q}$ is a bijective vertex labeling function that assigns to each vertex in $Q$ a label from $L_{V}^{Q}$. The label of a vertex $v \in V^Q_p$ is the variable; that of a vertex $v \in V^Q_l$ is its literal value; and that of a vertex $v \in V^Q_c \cup V^Q_e$ is its corresponding URI.
\item $f_{E}^{Q}:V^{Q} \rightarrow L_{E}^{Q}$ is a bijective vertex labeling function that assigns to each edge in $Q$ a label from $L_{V}^{Q}$.  An edge label can be a property or an edge variable.
\item $FL$ are constraint filters.
\end{enumerate}
\end{definition}

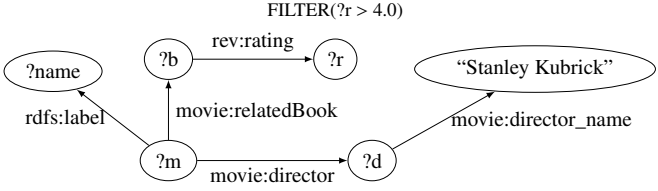
\begin{figure}
	\scalebox{0.8}{
\begin{tikzpicture}
	\node [style=ellipse,draw] (n1) {?m};
	\node [style=ellipse,draw] (n2) [right=2.5cm of n1] {?d} edge [latex-] node[below] {movie:director} (n1);
	\node [style=ellipse,draw] (n3) [above left=of n1] {?name} edge [latex-] node[left] {rdfs:label} (n1);
	\node [style=ellipse,draw] (n4) [above =of n1] {?b} edge [latex-] node[right] {movie:relatedBook} (n1);
	\node [style=ellipse,draw] (n5) [above right=of n2] {``Stanley Kubrick''} edge [latex-] node[right] {movie:director\_name} (n2);
	\node [style=ellipse,draw] (n6) [right=2cm of n4] {?r} edge [latex-] node[above] {rev:rating} (n4);
		\node[attribute] (n6a) [above=0.1cm of n6]{\small{FILTER(?r $>$ 4.0)}};
\end{tikzpicture}
	} 
	\caption{SPARQL query graph corresponding to query $Q_{1}$}
	\label{fig:query_graph}
\end{figure}

The query graph for $Q_{1}$ is given in Figure \ref{fig:query_graph}.

The semantics of SPARQL query evaluation can, therefore, be defined as subgraph matching using graph homomorphism whereby all subgraphs of an RDF graph $G$ are found that are homomorphic to the SPARQL query graph $Q$. In this context, OPT represents the optional triple patterns that may be matched.

\begin{definition}[SPARQL graph match]\label{def:match}
	Consider an RDF graph $G$ and a query graph $Q$ that has $n$ vertices $\{v_1,...,v_n\}$. A set of $n$ distinct vertices $\{u_1,...,u_n\}$ in $G$ is said to be a \emph{match} of $Q$, if and only if there exists a bijective function $F$, where $u_i=F(v_i)$ ($1 \leq i \leq n$) , such that :
\begin{enumerate}
\item If $v_i$ is a literal vertex, $v_i$ and $u_i$ have the same literal value;
\item If $v_i$ is an entity or class vertex, $v_i$ and $u_i$ have the same URI;
\item If $v_i$ is a variable vertex, $u_i$ should satisfy the filter constraint over parameter vertex $v_i$ if any; otherwise, there is no constraint over $u_i$;
\item If there is an edge from $v_i$ to $v_j$ in $Q$, there is also an edge from $u_i$ to $u_j$ in $G$. If the edge label in $Q$ is $p$ (i.e., property), the edge from $u_i$ to $u_j$ in $G$ has the same label. If the edge label in $Q$ is a parameter,
    the edge label should satisfy the corresponding filter constraint; otherwise, there is no constraint over the edge label from $u_i$ to $u_j$ in $G$.
\end{enumerate}
\end{definition}

It is usual to talk about SPARQL query types based on the shape of the query graph.  Typically, three query types are observed: (i) linear (Figure~\ref{fig:sample-sparql}a), where the variable in the object field of one triple pattern appears in the subject of another triple pattern (e.g., ?y in $Q_{1}$) (ii) star-shaped (Figure~\ref{fig:sample-sparql}b), where the variable in the object field of one triple pattern appears in the subject of multiple other triple patterns (e.g., ?a in $Q_{2}$), and (iii) snowflake-shaped (Figure~\ref{fig:sample-sparql}c), which is a combination of multiple star queries.

\begin{figure}
	\centering
	\begin{subfigure}[$Q_{1}$]{
			\scalebox{0.5}{\begin{tikzpicture}
	\coordinate (x1) at (0,0);
	\shiftAndRotate{x2}{x1}{3}{90}
	\shiftAndRotate{x3}{x2}{3}{90}

	\drawTriple{x1}{x2}{?x}{?y}{A}{left}
	\drawTriple{x2}{x3}{?y}{?z}{?b}{left}
\end{tikzpicture}
}
		}
	\end{subfigure}
	\begin{subfigure}[$Q_{2}$]{
			\scalebox{0.5}{\begin{tikzpicture}
	\coordinate (x1) at (0,0);
	\shiftAndRotate{x2}{x1}{3}{90}
	\shiftAndRotate{x3}{x2}{3}{120}
	\shiftAndRotate{x4}{x2}{3}{60}
	
	\drawTriple{x2}{x1}{?a}{?z}{C}{left}
	\drawTriple{x2}{x3}{?a}{?x}{A}{right}
	\drawTriple{x2}{x4}{?a}{?y}{B}{left}
\end{tikzpicture}
}
		}
	\end{subfigure}
	\begin{subfigure}[$Q_{3}$]{
			\scalebox{0.5}{\begin{tikzpicture}
	\coordinate (x1) at (0,0);
	\shiftAndRotate{x2}{x1}{3}{90}
	\shiftAndRotate{x3}{x2}{3}{120}
	\shiftAndRotate{x4}{x2}{3}{60}

	\shiftAndRotate{x5}{x4}{3}{120}
	\shiftAndRotate{x6}{x4}{3}{60}
	\shiftAndRotate{x7}{x4}{3}{300}

	\shiftAndRotate{x8}{x7}{3}{270}
	\shiftAndRotate{x9}{x7}{3}{60}

	\shiftAndRotate{x10}{x9}{3}{60}
	\shiftAndRotate{x11}{x9}{3}{300}
	
	\drawTriple{x2}{x1}{?f}{?g}{E}{left}
	\drawTriple{x2}{x3}{?f}{?h}{E}{right}
	\drawTriple{x2}{x4}{?f}{?x}{D}{left}

	\drawTriple{x4}{x5}{?x}{?b}{D}{right}
	\drawTriple{x4}{x6}{?x}{?c}{D}{left}
	\drawTriple{x7}{x4}{?a}{?x}{A}{right}

	\drawTriple{x7}{x8}{?a}{?z}{C}{left}
	\drawTriple{x7}{x9}{?a}{?y}{B}{left}

	\drawTriple{x9}{x10}{?y}{?d}{D}{left}
	\drawTriple{x9}{x11}{?y}{?e}{D}{left}
\end{tikzpicture}
}
		}
	\end{subfigure}
	\caption{Sample SPARQL queries}
	\label{fig:sample-sparql}
\end{figure}


\section{Data Warehousing Approaches}
\label{sec:warehouse}

In this section we consider the approaches that take a centralized approach where the entire data is maintained in one RDF database. These fall into five categories: those that map the RDF data directly into a relational system, those that use a relational schema with extensive indexing (and a native storage system), those that denormalize the triples table into clustered properties, those that use column-store organization, and those that exploit the native graph pattern matching semantics of SPARQL.

Many of the approaches compress the long character strings to integer values using some variation of dictionary encoding in order to avoid expensive string operations. Each string is mapped to an integer in a mapping table, and that integer is then used in the RDF triple table(s). This facilitates fast indexing and access to values, but involves a level of indirection through the mapping table to get at the original strings. Therefore, some of these systems (e.g., Jena  \cite{Wilkinson2006}) employ encoding only for strings that are longer than a threshold. We ignore encoding in this paper, for clarity of presentation, and represent the data in its original string form.

\subsection{Direct Relational Mappings}

RDF triples have a natural tabular structure. A direct approach to handle RDF data using relational databases is to create single table with three columns (Subject, Property, Object) that holds the triples (there usually are additional auxiliary tables, but we ignore them here). The SPARQL query can then be translated into SQL and executed on this table. It has been shown that SPARQL 1.0 can be full translated to SQL \cite{Angles:2008aa,Sequeda:2014aa}; whether the same is true for SPARQL 1.1 with its added features is still open. This approach aims to exploit the well-developed relational storage, query processing and optimization techniques in executing SPARQL queries. Systems such as Sesame SQL92SAIL\footnote{Sesame is built to interact with any storage system since it implements a Storage and Inference Layer (SAIL) to interface with the particular storage system on which it sits. SQL92SAIL is the specific instantiation to work on relational systems.} \cite{Broekstra2002} and Oracle \cite{Chong2005} follow this approach.

Assuming that the table given in Figure \ref{fig:rdftable} is a relational table, the example  SPARQL query given earlier can be translated to the following SQL query (where s,p,o correspond to column names: Subject, Property, Object ):

\begin{lstlisting}[language=SQL,basicstyle=\scriptsize,showstringspaces=false]
SELECT T1.object
FROM  T as T1, T as T2, T as T3, 
      T as T4, T as T5
WHERE T1.p="rdfs:label" 
AND T2.p="movie:relatedBook"
AND T3.p="movie:director"
AND T4.p="rev:rating"
AND T5.p="movie:director_name"
AND T1.s=T2.s 
AND T1.s=T3.s
AND T2.o=T4.s
AND T3.o=T5.s
AND T4.o > 4.0 
AND T5.o="Stanley Kubrick"
\end{lstlisting}

An immediate problem that can be observed with this approach is the high number of self-joins -- these are not easy to optimize.  Furthermore, in large data sets, this single triples table becomes very large, further complicating query processing. 

\subsection{Single Table Extensive Indexing}

One alternative to the problems created by direct relational mapping is to develop native storage systems that allow extensive indexing of the triple table. Hexastore \cite{vldb08_Weiss:2008} and RDF-3X \cite{vldb08_Neumann:2008,Neumann2009} are examples of this approach. The single table is  maintained, but extensively indexed. For example, RDF-3X creates indexes for all six possible permutations of the subject, property, and object: (spo, sop,ops,ops,sop,pos). Each of these indexes are sorted lexicographically by the first column, followed by the second column, followed by the third column. These are then stored in the leaf pages of a clustered B$^{+}$-tree.

The advantage of this type of organization is that SPARQL queries can be efficiently processed regardless of where the variables occur (subject, property, object) since one of the indexes will be applicable. Furthermore,  it allows for index-based query processing that eliminates some of the self-joins -- they are turned into range queries over the particular index. Even when joins are required, fast merge-join can be used since each index is sorted on the first column. The obvious disadvantages are, of course, the space usage, and the overhead of updating the multiple indexes if data is dynamic. 

\subsection{Property Tables}

Property tables approach exploits the regularity exhibited in RDF datasets where there are repeated occurrence of patterns of statements. Consequently, it stores ``related'' properties in the same table.  The first system that proposed this approach is Jena \cite{Wilkinson2006}; IBM's DB2RDF \cite{BorneaDKSDUB13} also follows the same strategy. In both of these cases, the resulting tables are mapped to a relational system and the queries are converted to SQL for execution.

\begin{figure}
\centering
\subfigure [] {

\begin{tikzpicture}
\node[rectangle split, draw,rectangle split horizontal, rectangle split parts=5,minimum height=1cm,align=center] {Subject \nodepart{two} Property \nodepart{three}Property \nodepart{four}$\ldots$ \nodepart{five}Property};
\end{tikzpicture}
} 

\subfigure [] {

\begin{tikzpicture}
\node[rectangle split, draw,rectangle split horizontal, rectangle split parts=2,minimum height=1cm,align=center] {Subject \nodepart{two} Property};
\end{tikzpicture}
}

\subfigure [] {

\begin{tikzpicture}
\node[rectangle split, draw,rectangle split horizontal, rectangle split parts=6,minimum height=1cm,align=center] {Subject \nodepart{two} Property \nodepart{three}Property \nodepart{four}$\ldots$ \nodepart{five}Property \nodepart{six}Type};
\end{tikzpicture}
} 

\caption{Clustered property table design}
\label{fig:proptable}
\end{figure}
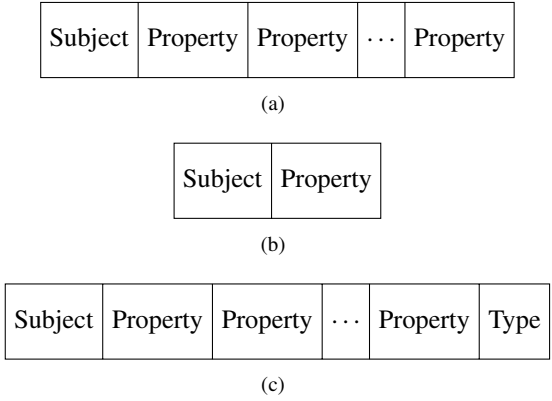

Jena defines two types of property tables. The first type, which can be called \emph{clustered property table},  group together the properties that tend to occur in the same (or similar) subjects. It defines different table structures for single-valued properties versus multi-valued properties. For single-valued properties, the table contains the subject column and a number of property columns (Figure \ref{fig:proptable}(a)). The value for a given property may be null if there is no RDF triple that uses the subject and that property. Each row of the table represents a number of RDF triples -- the same number as the non-null property values. For these tables, the subject is the primary key. For multi-valued properties, the table structure includes the subject and the multi-valued property (Figure \ref{fig:proptable}(b)). Each row of this table represents a single RDF triple; the key of the table is the compound key (subject,property). The mapping of the single triple table to property tables is a database design problem that is done by a database administrator.

Jena also defines a \emph{property class table} that cluster the subjects with the same \emph{type} of property into one property table (Figure~\ref{fig:proptable}(c)). In this case, all members of a class (recall our discussion of class structure within the context of RDFS) together in one table. The ``Type'' column is the value of  {\small \textsf{rdf:type}} for each property in that row.


The example dataset in Figure \ref{fig:rdftable} may be organized to create one table that includes the properties of subjects that are films, one table for properties of directors, one table for properties of actors, one table for properties of books and so on. Figure \ref{fig:example-property} shows one of these tables corresponding to the film subject. Note that the ``actor'' property is multi-valued (since there are two of them in film/2014), so a separate table is created for it.  

\begin{figure*}[t]
\centering
\footnotesize
\subfigure{
\begin{tabular}{|l|l|l|l|l|l|l|l|} \hline
Subject & label & initial\_release\_date & director & music\_contributor & based\_near & relatedBook & language \\ \hline
film/2014 & ``The Shining'' & ``1980-05-23'' & director/8476 & music\_contributor/4110 & 2635167 & 0743424425 & iso639-3/eng \\
film/2685 & ``A Clockwork Orange'' & NULL & director/8476 & NULL & NULL & NULL & NULL \\
film/424 & ``Spartakus'' & NULL & director/8476 & NULL & NULL & NULL & NULL \\
film/1267 & ``The Last Tycoon'' & NULL & NULL & NULL & NULL & NULL & NULL \\
film/3418 & ``The Passenger'' & NULL  & NULL & NULL & NULL & NULL & NULL \\
\hline
\end{tabular}
}
\subfigure{
\begin{tabular}{|l|l|} \hline
Subject & actor \\ \hline
film/2014 & actor/29704 \\
film/2014 & actor/30013 \\
film/1267 & actor/29704 \\
film/3418 & actor/29704 \\
\hline
\end{tabular}
}
\caption{Property table organization of subject ``mdb:film'' from the example dataset (prefixes are removed)}
\label{fig:example-property}
\end{figure*}

\begin{figure}
\centering
\subfigure [DPH] {

\begin{tikzpicture}
\node[rectangle split, draw,rectangle split horizontal, rectangle split parts=9,minimum height=1cm,align=center] {Subject \nodepart{two} Spill \nodepart{three}Prop$_{1}$ \nodepart{four} val$_{1}$ \nodepart{five}Prop$_{2}$ \nodepart{six}val$_{2}$ \nodepart{seven} $\ldots$ \nodepart{eight}Prop$_{k}$ \nodepart{nine} val$_{k}$};
\end{tikzpicture}
}

\subfigure [DS] {

\begin{tikzpicture}
\node[rectangle split, draw,rectangle split horizontal, rectangle split parts=2,minimum height=1cm,align=center] {$l\_id$ \nodepart{two} value};
\end{tikzpicture}
}

\caption{DB2RDF table design}
\label{fig:db2rdf}
\end{figure}
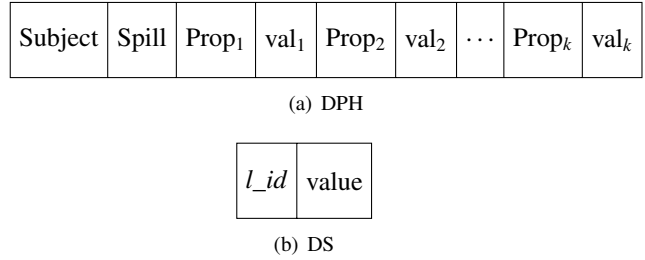

IBM DB2RDF \cite{BorneaDKSDUB13} also follows the same strategy, but with a more dynamic table organization (Figure \ref{fig:db2rdf}). The table, called \emph{direct primary hash} (DPH)  is organized by each subject, but instead of manually identifying ``similar'' properties, the table accommodates $k$ property columns, each of which can be assigned a different property in different rows. Each property column is, in fact, two columns: one that holds the property label, and the other that holds the value. If the number of properties for a given subject is greater than $k$, then the extra properties are spilled onto a second row and this is marked on the ``spill'' column. For multivalued properties, a \emph{direct secondary hash} (DSH) table is maintained -- the original property value stores a unique identifier $l\_id$, which  appears in the DS table along with the values. 

DB2RDF accomplishes the mapping from the single triples table into the DPH and DS tables automatically; the objective is to minimize the number of columns that are used in DPH while minimizing spills (since these cause expensive self-joins) that result from multiple properties being mapped to the same column. Note that, across all subjects, a property is always mapped to the same column; however, a given column can contain more than one property in different rows. The objective of the mapping  is to ensure that the columns can be overloaded with properties that do not occur together, but that properties that occur together are assigned to different columns. 


The advantage of property table approach is that joins in star queries (i.e., subject-subject joins) become single table scans. Therefore, the translated query has fewer joins.  The disadvantages are that in either of the two forms discussed above, there could be a significant number of null values in the tables (see the number of NULLs in Figure \ref{fig:example-property}), and dealing with multivalued properties requires special care. Furthermore, although star queries can be handled efficiently, this approach may not help much with other query types. Finally, when manual assignment is used, clustering ``similar'' properties is non-trivial and bad design decisions exacerbate the null value problem.

\subsection{Binary Tables}

Binary tables approach \cite{vldb07_Abadi:2007,Abadi:swstore} follows column-oriented database schema organization and defines a two-column table for each property containing the subject and object. This results in a set of tables each of which are ordered by the subject. This is a typical column-oriented database organization and benefits from the usual advantages of such systems such as reduced I/O due to reading only the needed properties and reduced tuple length, compression due to redundancy in the column values, etc. In addition, it avoids the null values that is experienced in property tables as well as the need for manual or automatic clustering algorithms for ``similar'' properties, and naturally supports multivalued properties -- each become a separate row as in the case of Jena's DS table. Furthermore, since tables are ordered on subjects, subject-subject joins can be implemented using efficient merge-joins. The shortcomings are that the queries require more join operations some of which may be subject-object joins that are not helped by the merge-join operation. Furthermore, insertions into the tables has higher overhead since multiple tables need to be updated. It has been argued that the insertion problem can be mitigated by batch insertions, but in dynamic RDF repositories the difficulty of insertions is likely to remain a significant problem. The proliferation of the number of tables  may have a negative impact on the scalability (with respect to the number of properties) of binary tables approach \cite{vldb08_Sidirourgos:2008}. 

For example, the binary table representation of the example dataset given in Figure \ref{fig:rdftable} would create one table for each unique property -- there are 18 of them. Two of these tables are shown in Figure \ref{fig:example-binary}.

\begin{figure}
\centering
\footnotesize
\subfigure[rdfs:label]{
\begin{tabular}{|l|l|} \hline
Subject & Object \\ \hline
film/2014 & ``The Shining'' \\
film/2685 & ``A Clockwork Orange'' \\
film/424 & ``Spartacus'' \\
film/1267 & ``The Last Tycoon'' \\
film/3418 & ``The Passenger'' \\
iso639-3/eng & ``English'' \\
\hline
\end{tabular}
}
\quad
\subfigure[movie:actor]{
\begin{tabular}{|l|l|} \hline
Subject & Object \\ \hline
film/2014 & actor/29704 \\
film/2014 & actor/30013 \\
film/1267 & actor/29704 \\
film/3418 & actor/29704 \\
\hline
\end{tabular}
}
\caption{Binary table organization of properties ``refs:label''  and ``movie:actor'' from the example dataset (prefixes are removed)}
\label{fig:example-binary}

\end{figure}

\subsection{Graph-based Processing}

Graph-based RDF processing approaches fundamentally implement the semantics of RDF queries as defined in Section \ref{sec:primer}. In other words, they maintain the graph structure of the RDF data (using some representation such as adjacency lists), convert the SPARQL query to a query graph, and do subgraph matching using homomorphism to evaluate the query against the RDF graph. Systems such as that proposed by B\"onstr\"om et al \cite{Bonstrom:2003aa}, gStore \cite{pvldb4_ZouMCOZ11,Zou:2013fk}, and chameleon-db \cite{Aluc:2015ab} follow this approach.

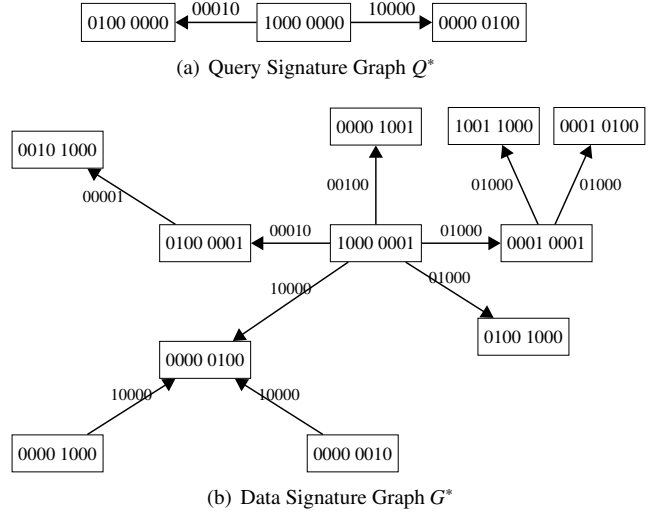
\begin{figure}
\centering

\subfigure [Query Signature Graph $Q^*$] {

\scalebox{0.72}{\begin{tikzpicture}[node distance=1.5cm, every edge/.style={link}]
\node[smentity] (ent1)  {0100 0000};
\node[smentity] (ent2) [right=of ent1] {1000 0000} edge[->] node[above] {00010} (ent1);
\node[smentity] (ent3) [right=of ent2] {0000 0100} edge[<-] node[above] {10000} (ent2);
\end{tikzpicture}  
}
}

\subfigure [Data Signature Graph $G^*$] {

\scalebox{0.7}{\begin{tikzpicture}[node distance=1.5cm, every edge/.style={link}]
\node[smentity,anchor=west] (001) {0010 1000};
\node[smentity,anchor=west] (002) [below right=of 001]{0100 0001} edge [->] node[left] {\small{00001}} (001);
\node[smentity,anchor=west] (003) [right=of 002]{1000 0001} edge [->] node[above] {\small{00010}} (002);
\node[smentity,anchor=west] (004) [below=of 002]{0000 0100} edge [<-] node[above] {\small{10000}} (003);
\node[smentity,anchor=west] (005) [below left=of 004]{0000 1000} edge [->] node[above] {\small{10000}} (004);
\node[smentity,anchor=west] (006) [below right=of 004]{0000 0010} edge [->] node[above] {\small{10000}} (004);
\node[smentity,anchor=west] (007) [above=of 003]{0000 1001} edge [<-] node[left] {\small 00100} (003);
\node[smentity,anchor=west] (010) [right=of 003]{0001 0001} edge [<-] node[above] {\small 01000} (003);
\node[smentity,anchor=west] (011) [below right=of 003]{0100 1000} edge [<-] node[above] {\small 01000} (003);
\node[smentity,anchor=west] (008) [above left=of 010,xshift=18mm,yshift=4.5mm]{1001 1000} edge [<-] node[left] {\small 01000} (010);
\node[smentity,anchor=west,xshift=-18mm,yshift=4.5mm] (008) [above right=of 010]{0001 0100} edge [<-] node[right] {\small 01000} (010);
\end{tikzpicture}  
}
}

\caption{Signature graphs}
\label{fig:signaturegraph}
\end{figure}

The advantage of this approach is that it maintains the original representation of the RDF data and enforces the intended semantics of SPARQL. The disadvantage is the cost of subgraph matching -- graph homomorphism is NP-complete. This raises issues with respect to the scalability of this approach to large RDF graphs; typical database techniques including indexing can be used to address this issue. In the remainder, we present the approach within the context of one system, gStore.

gStore is a graph-based triple store system that can answer different kinds of SPARQL queries -- exact queries, queries with wildcards (i.e., where partial information is known about a query object such as knowing the year of birth but not the full birthdate), and aggregate queries that are included in SPARQL 1.1 -- over dynamic RDF data repositories. It uses adjacency list representation of graphs. An important feature of gStore is to encode each  each entity and class vertex into a fixed length bit string. One motivation for this encoding is to deal with fixed length bit string rather than variable length character strings -- this is similar to the dictionary encoding mentioned above. The second, and more important, motivation is to capture the ``neighborhood'' information for each vertex in the encoding that can be exploited during graph matching. This results in the generation of a \emph{data signature graph} $G^*$, in which each vertex corresponds to a class or an entity vertex in the RDF graph. Specifically, $G^*$ is induced by all entity and class vertices in the original RDF graph $G$ together with the edges whose endpoints are either entity or class vertices. Figure \ref{fig:signaturegraph}(b) shows the data signature graph $G^*$ that corresponds to RDF graph $G$ in Figure \ref{fig:datagraph}. An incoming SPARQL query is also represented as a \emph{query graph} $Q$ that is similarly encoded into a \emph{query signature graph} $Q^*$. The encoding of query graph depicted in Figure \ref{fig:query_graph} into a query signature graph $Q^*_2$ is shown in Figure \ref{fig:signaturegraph}(a). 

The problem now turns into finding matches of $Q^*$ over $G^*$. Although both the RDF graph and the query graph are smaller as a result of encoding, the NP-completeness of the problem remains.  Therefore, gStore uses a filter-and-evaluate strategy to reduce the search space over which matching is applied. The objective is to first use a false-positive pruning strategy to find a set of candidate subgraphs (denoted as $CL$), and then  validate these using the adjacency list to find answers (denoted as $RS$). Accordingly, two issues need to be addressed. First, the encoding technique should guarantee that $RS \subseteq CL$ -- the encoding described above provably achieves this. Second, an efficient subgraph matching algorithm is required to find matches of $Q^*$ over $G^*$.  For this, gStore uses an index structure called VS$^*$-tree that is a summary graph of $G^*$. VS$^{*}$-tree is used to efficiently process queries using a pruning strategy to reduce the search space for finding matches of $Q^*$ over $G^*$.

\section{Distributed RDF Processing}
\label{sec:distributed}

In the previous section we focused on centralized, single machine approaches to RDF data management and SPARQL query processing. In this section, we focus on distributed approaches. This section is taken from \cite{Peng:2015aa}. We identify and discus four classes of approaches: cloud-based solutions, partitioning-based approaches, federated SPARQL evaluation systems, and partial evaluation-based approach.

\subsection{Cloud-based Approaches}

There have been a number of works (e.g., \cite{DBLP:JenaHBase,DBLP:SHARD,Husain:2011aa,Zhang:2012ab,ICDE13_Zhang,Zeng:2013aa,Papailiou:2012aa,Papailiou:2014aa}) focusing on managing large RDF datasets using existing cloud platforms; a very good survey of these is is provided by Saoudi and Manolescu \cite{Kaoudi:2014aa}. Many of these approaches follow the MapReduce paradigm; in particular they use HDFS, and  store RDF triples in flat files in HDFS. When a SPARQL query is issued,  the HDFS files are scanned to find the matches of each triple pattern, which are then joined  using one of the MapReduce join implementations (see \cite{Li:2013uq} for more detailed description of these). The most important difference among these approaches is how the RDF triples are stored in HDFS files; this determines how the triples are accessed and the number of MapReduce jobs. In particular, SHARD \cite{DBLP:SHARD} directly stores the data in a single file and each line of the file represents all triples associated with a distinct subject. HadoopRDF \cite{Husain:2011aa} and PredicateJoin \cite{Zhang:2012ab} further partition RDF triples based on the property and store each partition within one HDFS file. EAGRE \cite{ICDE13_Zhang} first groups all subjects with similar properties into an entity class, and then constructs a compressed RDF graph containing only entity classes and the connections between them. It partitions the compressed RDF graph  using the METIS algorithm \cite{Karypis:1995aa}. Entities are placed into HDFS according to the partition set that they belong to.

Besides the HDFS-based approaches, there are also some works  that use other NoSQL distributed data stores to manage RDF datasets. JenaHBase \cite{DBLP:JenaHBase} and H$_2$RDF \cite{Papailiou:2012aa,Papailiou:2014aa} use some permutations of subject, property, object to build indices that are then stored in HBase (http://hbase.apache.org). Trinity.RDF \cite{Zeng:2013aa} uses the distributed memory-cloud graph system Trinity \cite{ShaoWL13} to index and store the RDF graph. It uses hashing on the vertex values to obtain a disjoint partitioning of the RDF graph that is placed on nodes in a cluster.

These approaches benefit from the high scalability and fault-tolerance offered by cloud platforms, but may suffer lower performance due to the difficulties of adapting MapReduce to graph computation.

\subsection{Partitioning-based Approaches}

The partition-based approaches \cite{pvldb4_HuangAR11,DBLP:WARP,DBLP:Partout,pvldb6_Lee2013,Sigmod14_Gurajada:2014} divide an RDF graph $G$ into several fragments and place each at a different site in a parallel/distributed system. Each site hosts a centralized RDF store of some kind. At run time, a SPARQL query $Q$ is decomposed into several subqueries such that each subquery can be answered locally at one site, and the results are then agregated. Each of these papers proposes its own data partitioning strategy, and different partitioning strategies result in different query processing methods.

In GraphPartition \cite{pvldb4_HuangAR11}, an RDF graph $G$ is partitioned into $n$ fragments, and each fragment is extended by including $N$-hop neighbors of boundary vertices. According to the partitioning strategy, the diameter of the graph corresponding to each decomposed subquery should not be larger than $N$ to enable subquery processing at each local site. WARP \cite{DBLP:WARP} uses some frequent structures in workload to further extend the results of GraphPartition. Partout \cite{DBLP:Partout} extends the concepts of minterm predicates in relational database systems, and uses the results of minterm predicates as the fragmentation units. Lee et. al. \cite{pvldb6_Lee2013} define the partition unit as a vertex and its neighbors, which they call a ``vertex block''. The vertex blocks are distributed based on a set of heuristic rules. A query is partitioned into blocks that can be executed among all sites in parallel and without any communication. TriAD uses METIS \cite{Karypis:1995aa} to divide the RDF graph into many partitions and the number of result partitions is much more than the number of sites. Each result partition is considered as a unit and distributed among different sites. At each site, TriAD maintains six large, in-memory vectors of triples, which correspond to all SPO permutations of triples. Meanwhile, TriAD constructs a summary graph to maintain the partitioning information.

All of the above methods implement particular  partitioning and distribution strategies that align with their specific requirements. When there is freedom to partition and distribute the data, this works fine, but there are circumstances when partitioning and distribution may be influenced by other requirements. For example, in some applications, the RDF knowledge bases are partitioned according to topics (i.e., different domains), or are partitioned according to different data contributors; in other cases, there may be administrative constraints on the placement of data.  In these cases, these approaches may not have the freedom to partition the data as they require. Therefore, partition-tolerant SPARQL processing may be desirable.

\subsection{Federated Systems}

Federated queries run SPARQL queries over multiple SPARQL endpoints. A typical example is linked data, where different RDF repositories are interconnected, providing a \emph{virtually integrated distributed database}. Federated SPARQL query processing is a very different environment than what we target in this paper, but we discuss these systems for completeness. 

A common technique is to precompute metadata for each individual SPARQL endpoint. Based on the metadata, the original SPARQL query is decomposed into several subqueries, where each subquery is sent to its relevant SPARQL endpoints. The results of subqueries are then joined together to answer the original SPARQL query. In DARQ \cite{Quilitz2008}, the metadata is called \emph{service description} that describes which triple patterns (i.e., property) can be answered. In \cite{Harth:2010aa}, the metadata is called Q-Tree, which is a variant of RTree. Each leaf node in Q-Tree stores a set of source identifers, including one for each source of a triple approximated by the node. SPLENDID \cite{DBLP:SPLENDID} uses Vocabulary of Interlinked Datasets (VOID) as the metadata. HiBISCuS \cite{Saleem:2014aa} relies on ``capabilities'' to compute the metadata. For each source, HiBISCuS defines a set of capabilities which map the properties to their subject and object authorities. TopFed \cite{Saleem:2014ab} is a biological federated SPARQL query engine whose metadata comprises of an N3 specification file and a Tissue Source Site to Tumour (TSS-to-Tumour) hash table, which is devised based on the data distribution.

In contrast to these, FedX \cite{Schwarte:2011aa} does not require preprocessing, but sends  ``SPARQL ASK'' to collect the metadata on the fly. Based on the results of ``SPARQL ASK'' queries, it decomposes the query into subqueries and assign subqueries with relevant SPARQL endpoints.

Global query optimization in this context has also been studied. Most federated query engines employ existing optimizers, such as dynamic programming \cite{aea}, for optimizing the join order of local queries. Furthermore, DARQ \cite{Quilitz2008} and FedX \cite{Schwarte:2011aa} discuss the use of semijoins  to compute a join between intermediate results at the control site and SPARQL endpoints.

\subsection{Partial Query Evaluation Approaches}

Partial function evaluation is a well-known programming language strategy whose basic idea is the following: given a function $f(s,d)$, where $s$ is the known  input and $d$ is the yet unavailable input, the part of $f$'s computation that depends only on $s$ generates a partial answer. This has been used for distributed SPARQL processing in the Distributed gStore system \cite{Peng:2015aa}.

An RDF graph is partitioned using some graph partitioning algorithm such as METIS \cite{Karypis:1995aa} (the particular graph partitioning algorithm does not matter as the approach is oblivious to it) into vertex-disjoint fragments  (edges that cross fragments are replicated in source and target fragments).  Each site receives the full SPARQL query $Q$ and executes it on the local RDF graph fragment providing data parallel computation. In this particular setting, the partial evaluation strategy is applied as follows: each site $S_i$ treats fragment $F_i$ as the known input in the partial evaluation stage; the unavailable input is the rest of the graph ($\overbar{G} = G \setminus F_{i}$). 

There are two important issues to be addressed in this framework. The first is to compute the partial evaluation results at each site $S_i$ given a query graph $Q$ -- in other words, addressing the graph homomorphism of $Q$ of $F_{i}$; this is called the  \emph{local partial match} since it finds the matches internal to fragment $F_{i}$. Since ensuring edge disjointness is not possible in vertex-disjoint partitioning, , there will be \emph{crossing edges} between graph fragments. The second task is the assembly of these local partial matches to compute crossing matches. Two different assembly strategies are proposed: \emph{centralized assembly}, where all local partial matches are sent to a single site, and \emph{distributed assembly}, where the local partial matches are assembled at a number of sites in parallel.

\section{Querying Linked Data}
\label{sec:lod}

As noted earlier, a major reason for the development of RDF is to facilitate the semantic web. An important aspect of this enterprise is the Web of Linked Data (WLD) that connects multiple web data sets encoded in RDF. In one sense, this is the web data integration, and querying the WLD is an important challenge.

WLD uses RDF to build the semantic web by following four principles:

\begin{enumerate}
\item All web resources are locally identified by their URIs.
\item Information about web resources/entities are encoded as RDF triples. In other words, RDF is the semantic web data model.\
\item Connections among data sets are established by data links.
\item Sites that host RDF data need to be able to service HTTP requests to serve up linked resources.
\end{enumerate}

Our earlier reference was to Linked Open Data (LOD), which enforces a fifth requirement on WLD:

\begin{enumerate}
\item[5.] The linked data content should be open.
\end{enumerate}
Examples of LOD datasets include DBpedia and Freebase. The following formalizes Web of Linked Data, ignoring the openness requirement.

The starting point is a Linked Document (LD) that is a web document with embedded RDF triples that encode web resources. These web documents are possibly interconnected to get the graph structure.

\begin{definition}[Web of Linked Data] \cite{esws12_Hartig12} \label{def:WLD}
Given an set $\mathcal{D}$ of Linked Documents (LD), a Web of Linked data is a tuple $W=(D,adoc, data)$ where:
\begin{itemize}
\item $D \subseteq \mathcal{D}$,
\item $adoc$ is a partial mapping from URIs to $D$, and
\item $data$ is a total mapping from $D$ to finite sets of RDF triples.
\end{itemize}
\end{definition}

Each of these documents may contain RDF triples that form \emph{data links} to other documents, which is formalized as follows.

\begin{definition}[Data Link] \cite{esws12_Hartig12}
A Web of Linked Data $W= (D, adoc, data)$ (as defined in Definition \ref{def:WLD}) contains a data link from document $d \in D$ to document $d' \in D$ if there exists a URI $u$ such that
\begin{itemize}
\item $u$ is mentioned in an RDF triple $t \in data(d)$, and
\item $d' = adoc(u)$
\end{itemize}
\end{definition}

The semantics of SPARQL queries over the WLD becomes tricky. One possibility is to adopt \emph{full web semantics} that specifies the scope of evaluating a SPARQL query expression to be all linked data. There is no known (terminating) query execution algorithm that can guarantee result completeness under this semantics. The alternative is a family of \emph{reachability-based semantics} that define the scope of evaluating a SPARQL query  in terms of the documents that can be reached: given a set of seed URIs and a reachability condition, the scope is all data along the paths of the data links from the seeds and that satisfy the reachability condition. The family is defined by different reachability conditions. In this case, there are computationally feasible algorithms.

There are three approaches to SPARQL query execution over WLD \cite{Hartig:2013aa}: traversal-based, index-based, and hybrid. \emph{Traversal approaches} \cite{Hartig13,DBLP:conf/esws/LadwigT11} basically implement a reachability-based semantics: starting from seed URIs, they recursively discover relevant URIs by traversing specific data links at query execution runtime. For these algorithms, the selection of the seed URIs is critical for performance. The advantage of traversal approaches is their simplicity (to implement) since they do not need to maintain any data structures (such as indexes). The disadvantages are the latency of query execution since these algorithms ``browse'' web documents, and repeated data retrieval from each document introduces significant latency. They also have limited possibility for parallelization -- they can be parallelized to the same extent that crawling algorithms can.

The \emph{index-based approaches} use an index to determine relevant URIs, thereby reducing the number of linked documents that need to be accessed. A reasonable index key is triple patterns \cite{wwwj11_UmbrichHKHP11} in which case the ``relevant'' URIs for a given query are determined by accessing the index, and the query is evaluated over the data retrieved by accessing those URIs. In these approaches, data retrieval can be fully parallelized, which reduces the negative impact of data retrieval on query execution time. The disadvantages of the approach are the dependence on the index -- both in terms of the latency that index construction introduces and in terms of the restriction the index imposes on what can be selected -- and the freshness issues that result from the dynamicity of the web and the difficulty of keeping the index up-to-date.

\emph{Hybrid approaches} \cite{ISWC10_LadwigT10} perform a traversal-based execution using prioritized listing of URIs for look-up. The initial seeds come from a pre-populated index; new discovered URIs that are not in the index are ranked according to number of referring documents.

\section{Conclusions}
\label{sec:conclusion}

In this paper, we gave a high level overview of RDF data management, focusing on the various approaches that have been adopted. The discussion focused on centralized RDF data management (what is called ``data warehousing approach'' in this paper), distributed RDF systems, and querying over the LOD data. There are many additional works on RDF that are omitted in this paper. Most notably, works on formal semantics of SPARQL, reasoning over RDF data, data integration using RDF, streaming RDF processing, and SPARQL processing under dynamic workloads \cite{Aluc:2015ab} are topics that are not covered.

\section*{Acknowledgements}

This paper is based on a tutorial that was given at ICDE 2014 conference The full citation is as follows:

O. Hartig  and M. T.  \"Ozsu. Linked Data query processing. Proc. 30th International Conference on Data Engineering. 2014,  pages 1286 -- 1289.

This research is partially supported by Natural Sciences and Engineering Research Council (NSERC) of Canada.

I thank Olaf Hartig for reading a previous version of this article and providing comments that improved the paper. 
\bibliography{FCS-Ozsu}

\begin{thebibliography}{10}

\bibitem{Fabian:yagoWWW07}
Suchanek F~M, Kasneci G, Weikum G.
\newblock Yago: a core of semantic knowledge.
\newblock In: Proc. 16th Int. World Wide Web Conf.
\newblock 2007,  697--706

\bibitem{Bizer:dbpedia}
Bizer C, Lehmann J, Kobilarov G, Auer S, Becker C, Cyganiak R, Hellmann S.
\newblock Dbpedia - a crystallization point for the web of data.
\newblock J. Web Sem., 2009, 7(3): 154--165

\bibitem{SBP14}
Schmachtenberg M, Bizer C, Paulheim H.
\newblock Adoption of best data practices in different topical domains.
\newblock In: Proc. 13th Int. Semantic Web Conf.
\newblock 2014,  245--260

\bibitem{Zhang:2012aa}
Zhang Y, Duc P~M, Corcho O, Calbimonte J~P.
\newblock {SRBench: A }streaming {RDF/SPARQL} benchmark.
\newblock In: Proc. 11th Int. Semantic Web Conf.
\newblock 2012,  641--657

\bibitem{Zaveri:2012aa}
Zaveri A, Rula A, Maurino A, Pietrobon R, Lehmann J, Auer S.
\newblock Quality assessment for linked data: A survey.
\newblock Web Semantics J., 2012,  1--5

\bibitem{Tang:2015aa}
Tang N.
\newblock Big {RDF} data cleaning.
\newblock In: Proc. Workshops of 31st Int. Conf. on Data Engineering.
\newblock 2015,  77--79

\bibitem{Klyne:aa}
Klyne G, Carroll J~J, McBride B.
\newblock {RDF} 1.1 concepts and abstract syntax.
\newblock Available at http://www.w3.org/TR/rdf11-concepts/; last accessed 17
  November 2015

\bibitem{Harris:aa}
Harris S, Seaborne A.
\newblock {SPARQL} 1.1 query language.
\newblock Available at http://www.w3.org/TR/sparql11-query/; last accessed 17
  November 2015

\bibitem{Zou:2013fk}
Zou L, {\"O}zsu M~T, Chen L, Shen X, Huang R, Zhao D.
\newblock {gStore}: A graph-based {SPARQL} query engine.
\newblock VLDB J., 2014, 23(4): 565--590

\bibitem{HartigOzsu2014}
Hartig O, {\"O}zsu M~T.
\newblock Reachable subwebs for traversal-based query execution.
\newblock 2014,  541--546 (Companion Volume)

\bibitem{esws12_Hartig12}
Hartig O.
\newblock {SPARQL} for a web of linked data: Semantics and computability.
\newblock In: Proc. 9th Extended Semantic Web Conf.
\newblock 2012,  8--23

\bibitem{W3C:2006aa}
W3C .
\newblock {SPARQL} query language for {RDF -- Formal} definitions.
\newblock Accessible at
  http://www.w3.org/2001/sw/DataAccess/rq23/sparql-defns.html\#defn\_GroupGraphPattern,
  2006.
\newblock last accessed 21 December 2015

\bibitem{Wilkinson2006}
Wilkinson K.
\newblock Jena property table implementation.
\newblock Technical Report HPL-2006-140, HP Laboratories Palo Alto, October
  2006

\bibitem{Angles:2008aa}
Angles R, Gutierrez C.
\newblock The expressive power of {SPARQL}.
\newblock In: Proc. 7th Int. Semantic Web Conf.
\newblock 2008,  114--129

\bibitem{Sequeda:2014aa}
Sequeda J~F, Arenas M, Miranker D~P.
\newblock {OBDA:} query rewriting or materialization? in practice, both!
\newblock In: Proc. 13th Int. Semantic Web Conf.
\newblock 2014,  535---551

\bibitem{Broekstra2002}
Broekstra J, Kampman A, Harmelen v~F.
\newblock Sesame: A generic architecture for storing and querying {RDF} and
  {RDF} schema.
\newblock In: Proc. 1st Int. Semantic Web Conf.
\newblock 2002,  54--68

\bibitem{Chong2005}
Chong E, Das S, Eadon G, Srinivasan J.
\newblock An efficient {SQL}-based {RDF} querying scheme.
\newblock In: Proc. 31st Int. Conf. on Very Large Data Bases.
\newblock 2005,  1216--1227

\bibitem{vldb08_Weiss:2008}
Weiss C, Karras P, Bernstein A.
\newblock Hexastore: sextuple indexing for semantic web data management.
\newblock Proc. VLDB Endowment, 2008, 1(1): 1008--1019

\bibitem{vldb08_Neumann:2008}
Neumann T, Weikum G.
\newblock {RDF-3X:} a {RISC}-style engine for {RDF}.
\newblock Proc. VLDB Endowment, 2008, 1(1): 647--659

\bibitem{Neumann2009}
Neumann T, Weikum G.
\newblock The {RDF-3X} engine for scalable management of {RDF} data.
\newblock VLDB J., 2009, 19(1): 91--113

\bibitem{BorneaDKSDUB13}
Bornea M~A, Dolby J, Kementsietsidis A, Srinivas K, Dantressangle P, Udrea O,
  Bhattacharjee B.
\newblock Building an efficient {RDF} store over a relational database.
\newblock In: Proc. ACM SIGMOD Int. Conf. on Management of Data.
\newblock 2013,  121--132

\bibitem{vldb07_Abadi:2007}
Abadi D~J, Marcus A, Madden S~R, Hollenbach K.
\newblock Scalable semantic web data management using vertical partitioning.
\newblock In: Proc. 33rd Int. Conf. on Very Large Data Bases.
\newblock 2007,  411--422

\bibitem{Abadi:swstore}
Abadi D~J, Marcus A, Madden S, Hollenbach K.
\newblock {SW-Store}: a vertically partitioned {DBMS} for semantic web data
  management.
\newblock VLDB J., 2009, 18(2): 385--406

\bibitem{vldb08_Sidirourgos:2008}
Sidirourgos L, Goncalves R, Kersten M, Nes N, Manegold S.
\newblock Column-store support for {RDF} data management: not all swans are
  white.
\newblock Proc. VLDB Endowment, 2008, 1(2): 1553--1563

\bibitem{Bonstrom:2003aa}
B\"onstr\"om V, Hinze A, Schweppe H.
\newblock Storing {RDF} as a graph.
\newblock In: Proc. 1st Latin American Web Congress.
\newblock 2003,  27 -- 36

\bibitem{pvldb4_ZouMCOZ11}
Zou L, Mo~J, Chen L, {\"O}zsu M~T, Zhao D.
\newblock {gStore:} answering {SPARQL} queries via subgraph matching.
\newblock Proc. VLDB Endowment, 2011, 4(8): 482--493

\bibitem{Aluc:2015ab}
Alu\c{c} G.
\newblock Workload Matters: A Robust Approach to Physical {RDF} Database
  Design.
\newblock PhD thesis, University of Waterloo, 2015

\bibitem{Peng:2015aa}
Peng P, Zou L, \"Ozsu M~T, Chen L, Zhao D.
\newblock Processing {SPARQL} queries over distributed {RDF} graphs.
\newblock VLDB J., 2015.
\newblock Forthcoming.

\bibitem{DBLP:JenaHBase}
Khadilkar V, Kantarcioglu M, Thuraisingham B~M, Castagna P.
\newblock {Jena-HBase}: A distributed, scalable and efficient {RDF} triple
  store.
\newblock In: Proc. International Semantic Web Conference Posters {\&} Demos
  Track.
\newblock 2012

\bibitem{DBLP:SHARD}
Rohloff K, Schantz R~E.
\newblock High-performance, massively scalable distributed systems using the
  mapreduce software framework: the shard triple-store.
\newblock In: Proc. Int. Workshop on Programming Support Innovations for
  Emerging Distributed Applications.
\newblock 2010.
\newblock Article No. 4

\bibitem{Husain:2011aa}
Husain M~F, McGlothlin J, Masud M~M, Khan L~R, Thuraisingham B.
\newblock Heuristics-based query processing for large {RDF} graphs using cloud
  computing.
\newblock IEEE Trans. Knowl. and Data Eng., 2011, 23(9): 1312--1327

\bibitem{Zhang:2012ab}
Zhang X, Chen L, Wang M.
\newblock Towards efficient join processing over large {RDF} graph using
  mapreduce.
\newblock In: Proc. 24th Int. Conf. on Scientific and Statistical Database
  Management.
\newblock 2012,  250--259

\bibitem{ICDE13_Zhang}
Zhang X, Chen L, Tong Y, Wang M.
\newblock {EAGRE}: Towards scalable {I/O} efficient {SPARQL} query evaluation
  on the cloud.
\newblock In: Proc. 29th Int. Conf. on Data Engineering.
\newblock 2013,  565--576

\bibitem{Zeng:2013aa}
Zeng K, Yang J, Wang H, Shao B, Wang Z.
\newblock A distributed graph engine for web scale {RDF} data.
\newblock Proc. VLDB Endowment, 2013, 6(4): 265--276

\bibitem{Papailiou:2012aa}
Papailiou N, Konstantinou I, Tsoumakos D, Koziris N.
\newblock H\({}_{\mbox{2}}\){RDF}: adaptive query processing on {RDF} data in
  the cloud.
\newblock 2012,  397--400 (Companion Volume)

\bibitem{Papailiou:2014aa}
Papailiou N, Tsoumakos D, Konstantinou I, Karras P, Koziris N.
\newblock H\({}_{\mbox{2}}\){RDF+}: an efficient data management system for big
  {RDF} graphs.
\newblock In: Proc. ACM SIGMOD Int. Conf. on Management of Data.
\newblock 2014,  909--912

\bibitem{Kaoudi:2014aa}
Kaoudi Z, Manolescu I.
\newblock {RDF} in the clouds: A survey.
\newblock VLDB J., 2015, 24: 67--91

\bibitem{Li:2013uq}
Li~F, Ooi B~C, {\"O}zsu M~T, Wu~S.
\newblock Distributed data management using {MapReduce}.
\newblock ACM Comput. Surv., 2014, 46(3): Article No. 31

\bibitem{Karypis:1995aa}
Karypis G, Kumar V.
\newblock Analysis of multilevel graph partitioning.
\newblock 1995.
\newblock Article No. 29

\bibitem{ShaoWL13}
Shao B, Wang H, Li~Y.
\newblock Trinity: a distributed graph engine on a memory cloud.
\newblock In: Proc. ACM SIGMOD Int. Conf. on Management of Data.
\newblock 2013,  505--516

\bibitem{pvldb4_HuangAR11}
Huang J, Abadi D~J, Ren K.
\newblock Scalable {SPARQL} querying of large {RDF} graphs.
\newblock Proc. VLDB Endowment, 2011, 4(11): 1123--1134

\bibitem{DBLP:WARP}
Hose K, Schenkel R.
\newblock {WARP}: Workload-aware replication and partitioning for {RDF}.
\newblock In: Proc. Workshops of 29th Int. Conf. on Data Engineering.
\newblock 2013,  1--6

\bibitem{DBLP:Partout}
Galarraga L, Hose K, Schenkel R.
\newblock Partout: a distributed engine for efficient {RDF} processing.
\newblock 2014,  267--268 (Companion Volume)

\bibitem{pvldb6_Lee2013}
Lee K, Liu L.
\newblock Scaling queries over big rdf graphs with semantic hash partitioning.
\newblock Proc. VLDB Endowment, 2013, 6(14): 1894--1905

\bibitem{Sigmod14_Gurajada:2014}
Gurajada S, Seufert S, Miliaraki I, Theobald M.
\newblock {TriAD}: A distributed shared-nothing {RDF} engine based on
  asynchronous message passing.
\newblock In: Proc. ACM SIGMOD Int. Conf. on Management of Data.
\newblock 2014,  289--300

\bibitem{Quilitz2008}
Quilitz B.
\newblock Querying distributed {RDF} data sources with {SPARQL}.
\newblock In: Proc. 5th European Semantic Web Conf.
\newblock 2008,  524--538

\bibitem{Harth:2010aa}
Harth A, Hose K, Karnstedt M, Polleres A, Sattler K, Umbrich J.
\newblock Data summaries for on-demand queries over linked data.
\newblock In: Proc. 19th Int. World Wide Web Conf.
\newblock 2010,  411--420

\bibitem{DBLP:SPLENDID}
G{\"{o}}rlitz O, Staab S.
\newblock {SPLENDID: SPARQL} endpoint federation exploiting {VOID}
  descriptions.
\newblock In: Proc. ISWC 2011 Workshop on Consuming Linked Data.
\newblock 2011

\bibitem{Saleem:2014aa}
Saleem M, Ngomo A~N.
\newblock {HiBISCuS}: Hypergraph-based source selection for {SPARQL} endpoint
  federation.
\newblock In: Proc. 11th Extended Semantic Web Conf.
\newblock 2014,  176--191

\bibitem{Saleem:2014ab}
Saleem M, Padmanabhuni S~S, Ngomo A~N, Iqbal A, Almeida J~S, Decker S, Deus
  H~F.
\newblock {TopFed: TCGA} tailored federated query processing and linking to
  {LOD}.
\newblock J. Biomedical Semantics, 2014, 5: 47

\bibitem{Schwarte:2011aa}
Schwarte A, Haase P, Hose K, Schenkel R, Schmidt M.
\newblock {FedX}: Optimization techniques for federated query processing on
  linked data.
\newblock In: Proc. 10th Int. Semantic Web Conf.
\newblock 2011,  601--616

\bibitem{aea}
Astrahan M, Blasgen M, Chamberlin D, Eswaran K, Gray J, Griffiths P, King W,
  Lorie R, McJones P, Mehl J, Putzolu G, Traiger I, Wade B, Watson V.
\newblock System r: Relational approach to database management.
\newblock ACM Trans. Database Syst., 1976, 1(2): 97--137

\bibitem{Hartig:2013aa}
Hartig O.
\newblock An overview on execution strategies for linked data queries.
\newblock Datenbankspektrum, 2013, 13(2): 89--99

\bibitem{Hartig13}
Hartig O.
\newblock {SQUIN}: a traversal based query execution system for the web of
  linked data.
\newblock In: Proc. ACM SIGMOD Int. Conf. on Management of Data.
\newblock 2013,  1081--1084

\bibitem{DBLP:conf/esws/LadwigT11}
Ladwig G, Tran T.
\newblock {SIHJoin}: Querying remote and local linked data.
\newblock In: Proc. 8th Extended Semantic Web Conf.
\newblock 2011,  139--153

\bibitem{wwwj11_UmbrichHKHP11}
Umbrich J, Hose K, Karnstedt M, Harth A, Polleres A.
\newblock Comparing data summaries for processing live queries over linked
  data.
\newblock World Wide Web J., 2011, 14(5-6): 495--544

\bibitem{ISWC10_LadwigT10}
Ladwig G, Tran T.
\newblock Linked data query processing strategies.
\newblock In: Proc. 9th Int. Semantic Web Conf.
\newblock 2010,  453--469

\end{thebibliography}
\bibliographystyle{fcs}

\begin{minipage}{\columnwidth}
\Biography{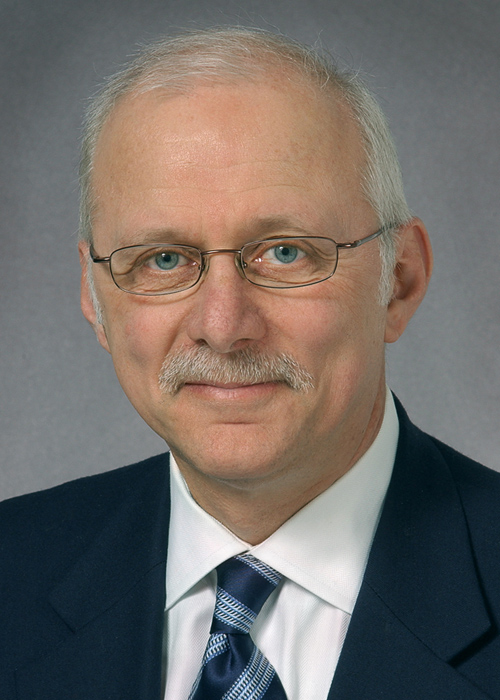}{M. Tamer \"Ozsu is Professor of Computer Science at the University of Waterloo. Dr. \"Ozsu's current research focuses on large scale data distribution, and management of unconventional data (e.g., graphs, RDF, XML, streams).

He is a Fellow of ACM and IEEE, an elected member of the Science Academy of Turkey, and a member of Sigma Xi and AAAS.}
\end{minipage}

\end{document}